\documentclass[onecolumn,showpacs,amssymb,prb,aps,superscriptaddress]{revtex4}
\usepackage{graphicx} 
\usepackage{epsfig}

\begin{document}
\raggedright

%----------------------------------------------------------------------
%               TITLE AND AUTHORS
%----------------------------------------------------------------------
\title{Thermal and nuclear quantum effects in the hydrogen bond dynamical symmetrization phase transition of $\delta$-AlOOH}
\author{Yael Bronstein} 
\author{Philippe Depondt}
\author{Fabio Finocchi} 
\affiliation{Sorbonne Universit\'es, UPMC Universit\'e Paris 6, UMR 7588, INSP, F-75005, Paris, France \\
CNRS, UMR 7588, INSP, F-75005, Paris, France}

%\maketitle

%----------------------------------------------------------------------
%                   ABSTRACT
%----------------------------------------------------------------------
\begin{abstract}

We conducted ab initio molecular dynamics simulations of the $\delta$
phase of the hydrous mineral AlOOH at ambient temperature and
high pressure. Nuclear quantum effects were included through a Langevin dynamics
in a bath of quantum harmonic oscillators. 
We confirm that under increasing pressure $\delta$-AlOOH
undergoes a phase transition from a $P2_1nm$ structure with asymmetric and disordered
O-H bonds to a stiffer $Pnnm$ phase with symmetric hydrogen bonds, which
should be stable within the pressure and temperature ranges typical
for the Earth's mantle.
The transition is initially triggered by proton tunneling, which makes
the mean proton position to coincide with the midpoint of the O-O
distance, at pressures as low as 10 GPa. However, only at much larger
pressures, around 30 GPa as previously found by other calculations,
the  $Pnnm$ phase with symmetric hydrogen bonds is stable from the
classical point of view. 
The transition is also characterized through the analysis of the H-O 
stretching modes, which soften considerably and fade out around 10 GPa 
in the $P2_1nm$ structure, when thermal and nuclear quantum effects are taken into
account in the simulations. At variance, the harmonic picture is not adequate to
describe the highly anharmonic effective potential that is seen by
the protons at the transition. 
Finally, we propose that the picture
of a dynamical transition to the high-symmetry and proton-centered $Pnnm$ phase, 
which is brought about by the onset of proton tunneling, could be confirmed by
quasi-elastic neutron scattering and vibrational spectroscopy under pressure.
\end{abstract}

%\pacs{62.50.p, 91.60.Hg, 64.60.Cn, 91.60.Ki}
%\date{}

\date{\today} 
\maketitle

%%%%%%%%%%%%%%%%%%%%%%%%%%%%%%%%%%%%%%%%%%%%%%%%%%%%%%%%%%%%%%%%%%%%%%
% --------------------------------------------------------------------
%                       MAIN TEXT
% --------------------------------------------------------------------
%%%%%%%%%%%%%%%%%%%%%%%%%%%%%%%%%%%%%%%%%%%%%%%%%%%%%%%%%%%%%%%%%%%%%%

%%%%%%%%%%%%%%%%%%%%%%%%%%%%%
%      INTRODUCTION         %
%%%%%%%%%%%%%%%%%%%%%%%%%%%%%
\section{Introduction}
Water transport in the Earth deep lower mantle is a cenral issue in geology geological studies \citep{ohtani2001}. In particular, hydrous minerals are believed to play a major role as water vectors
\citep{panero2004,sano2004,sano2008,ohira2014}.
Among the various dense hydrous minerals, aluminium oxide hydroxide (AlOOH) has received much attention recently, in particular its $\delta$ phase, which is a high-pressure polymorph of the more known diaspore ($\alpha$-AlOOH) and boehmite ($\gamma$-AlOOH). 
$\delta$-AlOOH has been synthesized for the first time in 2000 at $21$ GPa and $1000$ C \citep{suzuki2000}. 
Since then, it has been shown to be stable in a wide range of pressure and temperature \citep{ohtani2001,sano2004,sano2008}, making it a candidate as a hydrogen and water carrier within the deep mantle. 
Its structure is known through diffraction experiments
\citep{suzuki2000,ohtani2001,vanpeteghem2002,sano2004,sano2008,komatsu2006,kuribayashi2014} and theoretical calculations \citep{tsuchiya2002,panero2004}, except for the positions of the hydrogen atoms, which are not easily accessible experimentally. 
Moreover, isotope effects deriving from the differences between $\delta$-AlOOH and $\delta$-AlOOD \citep{sano-furukawa2009} indicate that nuclear quantum effects (NQE) play a role, as it happens in other minerals\citep{umemoto2015,hermann2016}.

Several studies suggested that the ionocovalent and hydrogen bonds in $\delta$-AlOOH become symmetric under increasing pressure, in contrast to $\alpha$-AlOOH \citep{winkler2001,friedrich2007}.
Recent synchrotron X-ray diffraction measurements indicated that there is a phase transition between $6.1$ and $8.2$ GPa, which is characterized by a modification of the compression properties of $\delta$-AlOOH \citep{kuribayashi2014}. 
However, this technique was unable to specify whether the post-transition phase consists in disordered and asymmetric hydrogen bonds, or if they are indeed symmetric. 

From the theoretical viewpoint, ab initio calculations predicted that $\delta$-AlOOH undergoes a second-order phase transition from an asymmetric (hydrogen off-centered - HOC) with group symmetry $P2_1nm$ to a symmetric (hydrogen centered - HC) structure with group symmetry $Pnnm$ at about $30$ GPa \citep{tsuchiya2002,tsuchiya2009}. 
Ab initio phonon dispersion curves computed at $0$ K showed that there are strong modifications of the modes involving protons at the transition \citep{tsuchiya2008}. 
However, the previous theoretical studies included neither thermal effects nor proton tunneling, which would be likely to reduce the transition pressure as in the case of the VII-X transition of ice under high pressure \citep{benoit1998,bronstein2014}.

However, the scenario for the transition in $\delta$-AlOOH seems to be more involved: in contrast with ice VII, it is not clear whether before the transition the protons are ordered or disordered \citep{komatsu2006,sano-furukawa2008,tsuchiya2008}; moreover, $\delta$-AlOOH belongs to the $P2_1nm$ space-group, a much lower symmetry than ice VII, with two inequivalent protons per unit cell in a less symmetric atomic environment. 
It is not a priori obvious that the same symmetric double-well picture stands for $\delta$-AlOOH as well. 
In particular, contrarily to the high-pressure ice structure, the oxygen-oxygen midpoint is not a symmetry center in  $P2_1nm$ $\delta$-AlOOH, whereas it is in the $Pnnm$ space-group; finally the experimental transition pressure \citep{kuribayashi2014}  ($\approx$ 10 GPa) is much lower than that predicted by static density functional theory calculations \citep{tsuchiya2008} ($\approx$ 30 GPa): dynamical effects in the symmetrization of the H-bonds may be relevant to characterize the transition at the atomistic scale.

We carry out, for the first time in $\delta$-AlOOH to our knowledge, ab initio molecular dynamics simulations at ambient temperature, in the $0$-$25$ GPa pressure range. The quantum nature of the nuclei is taken into account by coupling their dynamics to a "Quantum Thermal Bath" (QTB).\citep{dammak2009} 
From these simulations, we extract structural and dynamical quantities in order to investigate the hydrogen behavior in $\delta$-AlOOH and disentangle thermal from quantum effects as pressure increases.  
We anticipate that both thermal fluctuations and NQE play a crucial role in the modification of hydrogen bonds in $\delta$-AlOOH, contributing to a substantial downshift of the theoretical transition pressure, in agreement with the the experimental observations. 

After presenting in section \ref{sec:methods} the simulation techniques, including the QTB method, in section \ref{sec:structure}, we confront our simulation results with known structural data from mainly X-ray scattering. In section \ref{sec:spectra}, we compute vibrational properties that yield a clear signature of the phase transition at the transition pressure and discuss the link with experiments.
%Finally, in section \ref{sec:model}, a fully quantum one-dimensional analysis explains quite naturally the transition mechanism and disentangles NQE from classical thermal effects.

%%%%%%%%%%%%%%%%%%%%%%%%%%
%   SECTION 1 : METHODS  %
%%%%%%%%%%%%%%%%%%%%%%%%%%
\section{Methods}
\label{sec:methods}
%%%%%%%%%%%%%%%%%%%%%%%%%%
% SECTION 1.2 : QTB      %
%%%%%%%%%%%%%%%%%%%%%%%%%%
\subsection{The Quantum Thermal Bath}
For our simulations, we use the Quantum Espresso package \citep{giannozzi2009}, in which we implemented the QTB method. 
We perform molecular dynamics (MD) simulations in which a random force is added as well as a friction term according to a Langevin-type equation. When the nuclei are treated as classical particles (corresponding to standard MD), the random force simulates the thermal motion of the system \citep{kubo1978}, while in the QTB method, it simulates both thermal and quantum fluctuations \citep{callen1951,dammak2009}. 
The equations of motion for the nuclei are in both cases: 
\begin{equation}
m_i \frac{d^2 \mathbf{r}_i}{dt^2}(t) = \mathbf{f}_i(t) - m_i\gamma \frac{d\mathbf{r}_i}{dt}(t) + \mathbf{R}_i(t)
\end{equation}

\noindent where $\mathbf{r}_i$ is the position of a nucleus $i$ at time $t$, $m_i$ its mass and $\mathbf{f}_i$ the interatomic force applied on nucleus $i$ and computed via the Density Functional Theory (DFT). 
$\mathbf{R}_i$ is the random force, while $\gamma$ is a friction coefficient. 
The frequency distribution of the random force is related to the friction coefficient through its power spectrum $\left| \tilde{R}_i(\omega)\right|^2 = 2m_i \gamma \Theta(\omega,T)$ \citep{callen1951}, where $\tilde{R}(\omega)$ denotes the Fourier transform of $R(t)$ and the $\Theta$ function characterizes the kind of bath, either classical (standard Langevin MD) or quantum (QTB MD):

\begin{eqnarray}
\Theta(\omega,T) &=& \quad\quad k_B T \quad\quad \text{(standard Langevin MD)}  \\
\Theta(\omega,T) &=& \hbar \omega \left[ \frac{1}{2} + \frac{1}{\mathrm{exp}\left(\frac{\hbar \omega}{k_B T}\right) - 1} \right] \quad \text{(QTB MD)} 
\label{eq:qtb}
\end{eqnarray}

In practice, the only difference between standard Langevin MD and QTB MD is that in the standard case, $\mathbf{R}_i(t)$ has a Gaussian distribution with standard deviation $\sqrt{2m_i \gamma k_B T}$ ($T$ is the temperature and $k_B$ the Boltzmann constant), while in the QTB case, the random force is a colored noise. 
Therefore, from the comparison between standard Langevin MD and QTB trajectories, we are able to disentangle thermal from purely quantum effects. 
From now on, we denote with "standard Langevin MD" classical Langevin ab initio (DFT-based) molecular dynamics (thus using a white noise), while "QTB MD" denotes Langevin ab initio molecular dynamics using the colored noise in Equation \ref{eq:qtb} and thus including NQE. 

The QTB is exact for a system of harmonic oscillators, but also provides a good approximation for anharmonic systems \citep{dammak2012,calvo2012,savin2012,bronstein2014,bronstein2016}.
It has recently been the object of many studies that partly assessed its advantages and drawbacks \citep{calvo2012,basire2013,bedoya2014,hernandez2015,brieuc2016}.
In systems with few degrees of freedom, the probability distributions obtained via QTB capture most of the relevant quantum effects. Indeed, the computed probabilities along the atomic trajectories within the QTB approach are good approximations of the corresponding quantum distributions \citep{calvo2012,dammak2012,bronstein2014}, although a perfect quantitative agreement with exact methods can only be reached for harmonic solids \citep{dammak2009,basire2013,brieuc2016}.

With respect to other methods that include NQE such as path-integral-based molecular dynamics \citep{berne1986,marx1996}, the QTB has the advantage of presenting no additional computational cost compared to standard MD and to give access to vibrational information about the system, which is particularly suited for the present case and for more complex mineral structures. 
Vibrational spectra computed from QTB compare well with experiments and exact results both in molecules, clusters \citep{calvo2012}, and ice under pressure \citep{bronstein2014,bronstein2016}. 
Indeed, the spectra computed from Langevin MD, whether with a classical or a quantum noise, are a convolution of the real physical spectra with a Lorentzian of width $\gamma$, the friction coefficient \citep{izaguirre2001}. As long as $\gamma$ is chosen small enough, the vibrational spectra are reliable: hence, we set $\gamma=1$ THz, which is well below significant O-H frequencies in our system.

%%%%%%%%%%%%%%%%%%%%%%%%%%%%%%%%%%%
% SECTION 1.2 : TECHNICAL DETAILS %
%%%%%%%%%%%%%%%%%%%%%%%%%%%%%%%%%%%
\subsection{Technical details}
The random force is generated at the beginning of the simulation using a standard numerical technique to generate a noise with a prescribed correlation function \citep{maradudin1990}. 
In particular, no knowledge of the eigenfrequencies of the system is required before the simulation. 
The interatomic forces $\mathbf{f}_i$ are computed via the DFT within the generalized gradient approximation (GGA) \citep{perdew1996}, through the Becke-Lee-Yang-Parr functional \citep{becke1988,lee1988}. 
The interaction between the ionic cores and the valence electrons is described through ultra-soft pseudopotentials with non-linear core corrections. 
The Kohn-Sham orbitals are expanded in plane-waves with $50$ Ry energy cutoff. Our simulation cell is a $2\times 2\times 3$ supercell containing $24$ AlOOH units. 
The corresponding Brillouin Zone is sampled by a $2\times 2\times 2$ Monkhorst-Pack grid with a $1\times 1 \times 1$ offset (half a grid step in each direction). 
We run QTB and standard Langevin MD simulations at ambient temperature ($T = 300$ K) for several volumes. The time length of each simulation is about $29$ ps with a $0.484$ fs integration time step. 
The instantaneous pressures are calculated via the stress theorem \citep{nielsen1985}. From the atomic trajectories, we compute the mean pressures, and extract probability distributions and vibrational spectra.

%%%%%%%%%%%%%%%%%%%%%%%%%%%%%%%%%%%%%
% SECTION 2 : STRUCTURAL PROPERTIES %
%%%%%%%%%%%%%%%%%%%%%%%%%%%%%%%%%%%%%
\section{Structural properties}
\label{sec:structure}
%%%%%%%%%%%%%%%%%%%%%%%%%%%%%%%%%%%%%%
% SECTION 2.1 : P2_1nm vs $Pnnm$       %
%%%%%%%%%%%%%%%%%%%%%%%%%%%%%%%%%%%%%%

\subsection{$P2_1nm$ versus $Pnnm$ structures in $\delta-\mathrm{AlOOH}$}
First, we address from a static point of view and with no NQE, the apparent contradiction between theoretical \citep{tsuchiya2002,tsuchiya2009} computations that predict a $P2_1nm$ structure for pressures up to 30 GPa and X-ray experiments \citep{kuribayashi2014} which yield a $Pnnm$ symmetry above 10 GPa.

\begin{table}[h!] 
  \begin{tabular}{l|ccc|ccc|ccc} % Low pressure
    \multicolumn{10}{c}
                {Static lattice parameter optimization and experimental data}\\
                \hline
    & \multicolumn{3}{|c|}{$P2_1nm$-HOC; Theory}& 
    \multicolumn{3}{|c|}{$Pnnm$-HC; Theory}&
    \multicolumn{3}{|c}{$P2_1nm$-HOC; Exp.}\\
  & \multicolumn{3}{|c}{this work} &  
    \multicolumn{3}{|c}{this work} &  
    \multicolumn{3}{|c}{\cite{kuribayashi2014}} \\ \hline\hline
    Pressure (GPa) &\multicolumn{3}{|c|}{0} & \multicolumn{3}{|c|}{0} &
    \multicolumn{3}{|c}{0.0001} \\
    \hline
    & $x$ & $y$ & $z$ & $x$ & $y$ & $z$ & $x$ & $y$ & $z$  \\ %\hline
    \begin{tabular}{c}lattice\\parameters\end{tabular} ({\AA})
                       & 4.7882 & 4.2833 & 2.8617
                       & 4.7426 & 4.1944 & 2.8659
                       & 4.7127 & 4.2250 & 2.8310  \\
 \hline
 \rule{-1ex}{3ex}
 Al &\bf 0   &\bf 0.029 &\bf 0           &\bf 0   &\bf 0   &\bf 0      & \bf 0 &\bf 0.02433 &\bf 0  \\
 Al & 1/2 & 0.471 & 1/2         & 1/2 & 1/2 & 1/2    & 1/2 & 0.47567 & 1/2\\
 O$^{(1)}$ &\bf 0.341 &\bf 0.250 &\bf 0   &\bf 0.355 &\bf 0.240 &\bf 0  &\bf 0.34246 &\bf 0.24860 &\bf 0\\
 O$^{(1)}$ & 0.841 & 0.250 & 1/2  & 0.855 & 0.260 & 1/2& 0.84246 & 0.25140 & 1/2\\
 O$^{(2)}$ &\bf 0.642 &\bf 0.747 &\bf 0    & 0.645 & 0.760 & 0  &  \bf 0.6458 & \bf 0.74918 & \bf 0\\
 O$^{(2)}$ & 0.142 & 0.753 & 1/2  & 0.145 & 0.740 & 1/2&  0.1458 & 0.75082 & 1/2\\
 H &\bf 0.516 &\bf -0.059 &\bf 0         &\bf 1/2 &\bf 0   &\bf 0      & \\
 H & 0.016 &  0.559 & 1/2       & 0   & 1/2 & 1/2    & \\
 \hline \hline \hline % High pressure P=10GPa
   & \multicolumn{3}{|c|}{$P2_1nm$-HOC; Theory}&  
    \multicolumn{3}{|c|}{$Pnnm$-HC; Theory}&
    \multicolumn{3}{|c}{$Pnnm$-HC; Exp.}\\
  & \multicolumn{3}{|c}{this work} &  
    \multicolumn{3}{|c}{this work} &  
    \multicolumn{3}{|c}{\cite{kuribayashi2014}} \\ \hline\hline
    Pressure (GPa) &\multicolumn{3}{|c|}{10} & \multicolumn{3}{|c|}{10} &
    \multicolumn{3}{|c}{8.2} \\
    \hline
    & $x$ & $y$ & $z$ & $x$ & $y$ & $z$ & $x$ & $y$ & $z$  \\ %\hline
    \begin{tabular}{c}lattice\\parameters\end{tabular} ({\AA})
                       & 4.6829 & 4.1835 & 2.8201
                       & 4.6658 & 4.1404 & 2.8211
                       & 4.6379 & 4.1342 & 2.7990 \\
 \hline
 \rule{-1ex}{3ex}
 Al &\bf 0   &\bf 0.020 &\bf 0    &\bf 0   &\bf 0   &\bf 0     & \bf 0 &\bf 0 &\bf 0  \\
 Al & 1/2 & 0.480 & 1/2         & 1/2 & 1/2 & 1/2    & 1/2 & 1/2 & 1/2\\
 O$^{(1)}$ &\bf 0.343 &\bf 0.246 &\bf 0  &\bf 0.353 &\bf 0.240 &\bf 0  &\bf 0.3510 &\bf 0.2431 &\bf 0\\
 O$^{(1)}$ & 0.843 & 0.260 & 1/2  & 0.853 & 0.260 & 1/2& 0.8510 & 0.2569 & 1/2\\
 O$^{(2)}$ &\bf 0.645 &\bf 0.753 &\bf 0    & 0.647 & 0.760 & 0  &  0.6490 & 0.7369 & 0\\
 O$^{(2)}$ & 0.145 & 0.747 & 1/2  & 0.147 & 0.740 & 1/2&  0.1490 & 0.7531 & 1/2\\
 H &\bf 0.513 &\bf -0.042 &\bf 0         &\bf 1/2 &\bf 0   &\bf 0      & \\
 H & 0.013 &  0.542 & 1/2       & 0   & 1/2 & 1/2    & \\
% \end{tabular}
 
 \hline \hline \hline % High pressure P=30GPa
   & \multicolumn{3}{|c|}{$P2_1nm$-HOC; Theory}&  
    \multicolumn{3}{|c|}{$Pnnm$-HC; Theory}& \\
  & \multicolumn{3}{|c}{this work} &  
    \multicolumn{3}{|c|}{this work} &  \\
    \hline\hline
    Pressure (GPa) &\multicolumn{3}{|c|}{30} & \multicolumn{3}{|c|}{30} & \\
    \hline
    & $x$ & $y$ & $z$ & $x$ & $y$ & $z$ &   \\ %\hline
    \begin{tabular}{c}lattice\\parameters\end{tabular} ({\AA})
                       & 4.5557 & 4.0641 & 2.7475
                       & 4.5527 & 4.0572 & 2.7484
                        \\
 \hline
 \rule{-1ex}{3ex}
 Al        &\bf 0     &\bf 0.008 &\bf 0  &\bf 0     &\bf 0     &\bf  0 &  &  & \\
 Al        & 1/2      & 0.492    & 1/2   & 1/2      & 1/2      & 1/2   &  &  & \\
 O$^{(1)}$ &\bf 0.347 &\bf 0.242 &\bf 0  &\bf 0.350 &\bf 0.240 &\bf 0  &  &  & \\
 O$^{(1)}$ & 0.847    & 0.258    & 1/2   & 0.850    & 0.260    & 1/2   &  &  & \\
 O$^{(2)}$ &\bf 0.648 &\bf 0.759 &\bf 0  & 0.650    & 0.760    & 0     &  &  & \\
 O$^{(2)}$ & 0.148    & 0.741    & 1/2   & 0.150    & 0.740    & 1/2   &  &  & \\
 H         &\bf 0.506 &\bf -0.017 &\bf 0 &\bf 1/2   &\bf 0     &\bf 0 &  &  & \\
 H         & 0.006    &  0.547    & 1/2  & 0        &  1/2     & 1/2   &  &  & \\
 \end{tabular}

  \centering
  \caption{Equilibrium lattice parameters (\AA) and atomic positions in units of
    lattice vectors of $\delta$-AlOOH, computed within GGA-BLYP, through static
    optimization at 3 pressures($P=0$ (top), 10 (middle) and 30 GPa (bottom), compared with available experimental data from \cite{kuribayashi2014}.The irreducible atomic positions are in boldface characters.}
  \label{tab:table1}
  \end{table}

The relation between the atomic positions in $P2_1nm$ and $Pnnm$ space groups is not completely straightforward. From direct inspection to the International Tables of Crystallography, the two sets of positions can be continuously changed into one another, by shifting the origin of the atomic positions, all of the $(2a)$ kind, in $P2_1nm$ (space group 31) by $\delta y=-1/4$  with respect to the conventional origin.

After the $\delta y=-1/4$ shift, one can easily see that the $P2_1nm$ symmetry group can be continously tranformed into the $Pnnm$ one (see \emph{e.g.} table \ref{tab:table1}: atomic positions from static structure optimization and corresponding experimental data). 
More specifically, concerning the proton positions which are not easily detectable by X-ray scattering and subject to relevant quantum effects such as tunneling, we denote two kinds of structures, as previously done in the literature \citep{tsuchiya2002,tsuchiya2009}: 
off-centered Hydrogen (HOC) and centered Hydrogen (HC). 
While Hydrogen in $Pnnm$ can only exist as HC, one can build two $P2_1nm$ virtual structures: HOC and HC. 
In the former, the atomic positions are $(2a)$, while in the latter the Hydrogen positions are identical to the ones in $Pnnm$. 
Nevertheless, as a result of the loss of inversion symmetry, protons in $P2_1nm$-HC are not at the same distance from the two inequivalent O$^{(1)}$ and O$^{(2)}$ and their energy landscape is not symmetric in the order parameter $\xi=d_{\mathrm O^{(1)}-H} - d_{\mathrm O^{(2)}-H}$. 
For instance, for $P=0$ GPa (see table \ref{tab:table1}, top) the computed
$d_{\mathrm O-H}$ distances are: 1.03 {\AA} and 1.57 {\AA} ($P2_1nm$-HOC); 
 1.22 {\AA} and 1.22 {\AA} ($Pnnm$-HC).

  At this stage, table \ref{tab:table1} confirms with our static optimization data the previous ab-initio results by \cite{tsuchiya2002} and \cite{tsuchiya2009} and obtain quantitative agreement with experimental crystallographic data by \cite{kuribayashi2014}. In addition, consecutive dynamical matrix calculations provide several negative eigenvalues and thus imaginary frequencies for the $Pnnm$-HC structure (figure \ref{schema}) for $P<30$ GPa confirming that the HC structure is unstable up to that pressure in contradiction with experimental data: this means that static ab-initio optimization is unable to account for the experimentally obtained transition pressure and we can suspect that thermal effects and/or nuclear quantum effects play a significant role in this transition. Furthermore, that the two structures can be transformed continuously into one another, as also shown for the high pressure data at 30 GPa in table \ref{tab:table1}, indeed allows dynamical effects to play a role in the phase transition.

%%%%%%%%%%%%%%%%%%%%%%%%%%%%%%%%%%%%%
%  EQUATION OF STATE  %
%%%%%%%%%%%%%%%%%%%%%%%%%%%%%%%%%%%%%
\subsection{Equation of state}
\label{sec:eos}
\begin{figure}[tbp!]
\centering
\includegraphics[width=\linewidth]{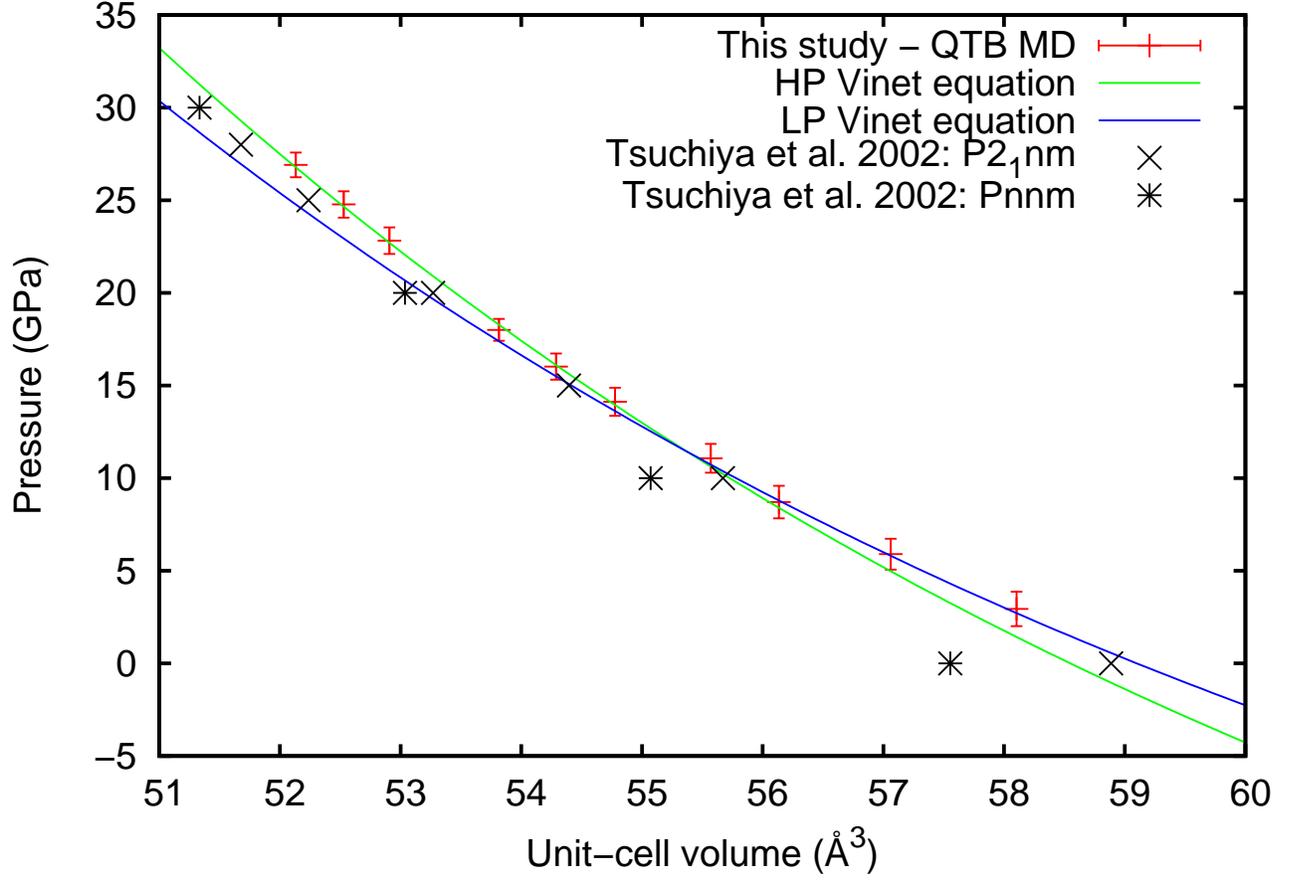}
\caption{Equation of state for $\delta$-AlOOH. The points computed through our molecular dynamics simulations at $T = 300$ K with the Quantum Thermal Bath are in red. The fitted curves to the Vinet equation\citep{vinet1987} in the low-pressure (LP) and the high-pressure (HP) structures are drawn in blue and green, respectively; the corresponding values for the equilibrium volumes, the bulk moduli and their derivatives are: $V_{LP} = 59.1\,\AA^3$, $V_{HP} = 58.5\,\AA^3$, $B_{LP} = 154.2$ GPa, $B_{HP} = 183.4$ GPa and $B_{LP}^\prime = B_{HP}^\prime = 4$. Also shown are the $(P,V)$ values computed by Tsuchiya and coworkers at $T = 0$ K in the $P2_1nm$ and the $Pnnm$ structures.\citep{tsuchiya2002}}
\label{eos}
\end{figure}

Synchrotron X-ray experiments showed a change of compressibility of $\delta$-AlOOH around $10$ GPa \citep{sano-furukawa2009}. 
Moreover, theoretical calculations conducted in the asymmetric ($P2_1nm$ - HOC) and symmetric ($Pnnm$ - HC) phases yielded two distinct bulk moduli \citep{tsuchiya2002,tsuchiya2009}. 
In the asymmetric $P2_1nm$ structure, up to $25$ GPa, our results from standard MD runs agree with those from \cite{tsuchiya2002}. 
The same procedure was carried out within the QTB frame: we fitted our computed $(P,V)$ points at $T = 300$ K through two distinct Vinet equations of state \citep{vinet1987}, one adapted for the low-pressure (LP) structure, the other for the high-pressure (HP) structure (see Figure \ref{eos}), as done in previous studies \citep{tsuchiya2002,sano-furukawa2009}. 
Within this procedure, the two equations of state cross between $10$ and $15$ GPa; accordingly, the bulk modulus of $\delta$-AlOOH changes from $154$ ($\pm 2$) GPa at low pressure to $183$ ($\pm 2$) GPa at high pressure (in both cases, $B' = 4$). 
The small discrepancies with the experiments \citep{sano-furukawa2009} are mainly due to the GGA functional, which slightly overestimates the cell parameters with respect to the experimental values. 
Our dynamical simulations are therefore consistent with a change of compressibility in $\delta$-AlOOH with an increase of the bulk modulus by about $20\%$ between $10$ and $15$ GPa at $T = 300$~K, which is compatible with a phase transition towards stiffer O-H-O bonds as indicated by experiments.
However, given the statistical error bars, we cannot demonstrate that a single equation of state cannot fit our computed points and we therefore resort to more precise methods, such as the analysis of the interatomic distances and the vibrational spectra, to establish the transition pressure in the following.

%%%%%%%%%%%%%%%%%%%%%%%%%%%%%%%%%%%%%%%
% INTERATOMIC DISTANCES %
%%%%%%%%%%%%%%%%%%%%%%%%%%%%%%%%%%%%%%%
\subsection{Interatomic distances}
\label{sec:pcf}
\begin{figure}[tbp!]
\centering
\includegraphics[width=\linewidth]{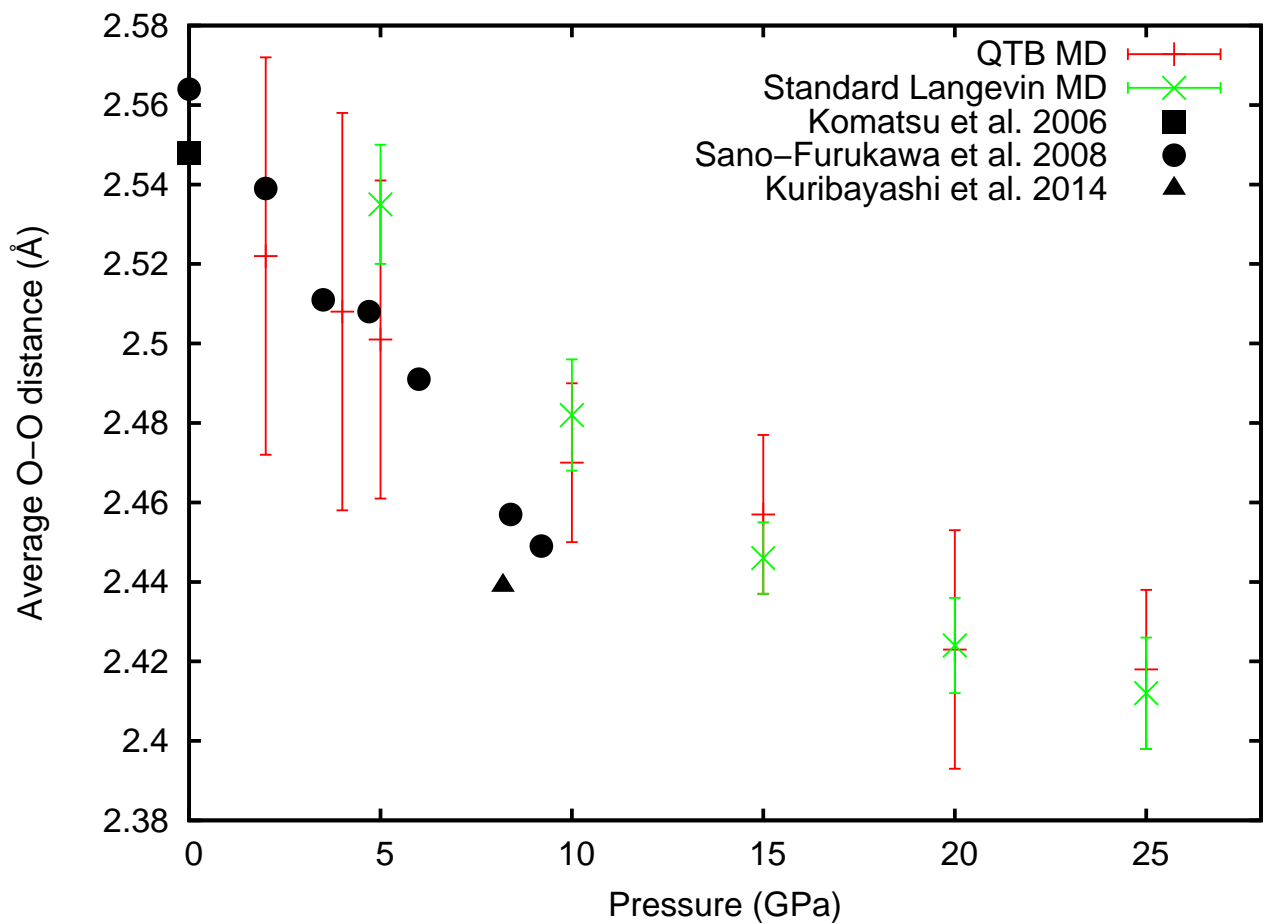}
\caption{Average oxygen-oxygen distance computed from standard Langevin molecular dynamics and Quantum Thermal Bath simulations. The vertical bars indicate the widths of the probability distributions of O-O. The black symbols indicate experimental results.\citep{komatsu2006,sano-furukawa2008,kuribayashi2014}}
\label{oo_dis}
\end{figure}

The O-O average distance is a relevant parameter for any transition from asymmetric O-H$\cdots$O to symmetric O-H-O bonds; indeed, in pure ice, proton tunneling and the onset of symmetrization happen at O-O distances around $2.4$ \AA\, \citep{benoit2005}, and this is generally accepted as a good criterion for the onset of symmetrization. 
The average O-O distances that were computed along standard Langevin and QTB simulations at $T = 300$~K are reported in Figure \ref{oo_dis}. 
They are very similar, apart from larger fluctuations within the QTB framework, which are mainly due to zero-point motion. 
Our results are in agreement with experimental data \citep{komatsu2006,sano-furukawa2008,kuribayashi2014}, though we slightly overestimate the O-O distance as pressure increases to about $8$ GPa with respect to very recent X-ray data \citep{kuribayashi2014}. One may note that both experimental results and ours for the O-O distance, remain significantly above the 2.4 \AA\, threshold until over 25 GPa. 

\begin{figure}[tbp!]
\centering
\includegraphics[width=\linewidth]{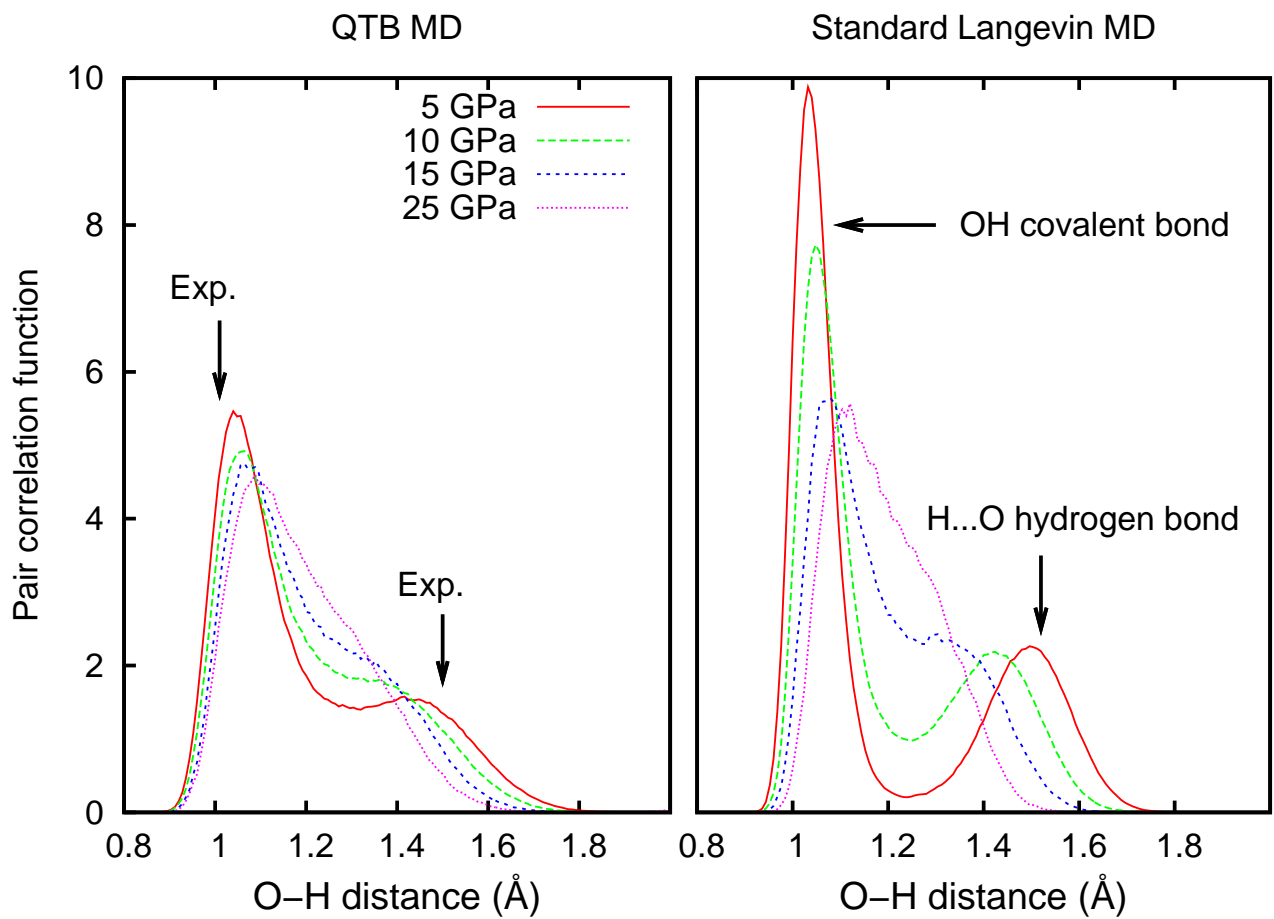}
\caption{Oxygen-hydrogen pair correlation function for different pressures between $5$ and $25$ GPa, obtained from Quantum Thermal Bath (left) and standard Langevin (right) molecular dynamics simulations at ambient temperature. The two peaks indicate two different distances: the ionocovalent O-H bond length at approximately $1.1$ \AA\, and the H$\cdots$O hydrogen bond around $1.5$ \AA. The two arrows in the left panel indicate neutron scattering measurements on $\delta$-AlOOD at $5$ GPa.\citep{sano-furukawa2008}}
\label{pcf}
\end{figure}

In order to investigate the hydrogen bonds in $\delta$-AlOOH, we compute the oxygen-hydrogen (O-H) pair correlation function (PCF), averaged over the trajectories obtained from QTB or standard Langevin MD (see Figure \ref{pcf}). 
At low pressure, the PCF displays two peaks indicating two distinct bond lengths: around $1.05$ \AA\, for ionocovalent O-H bonds and around $1.5$ \AA\, for hydrogen H$\cdots$O bonds. 
These distances are in agreement with experimental measurements on $\delta$-AlOOD obtained through neutron scattering around $5$ GPa \citep{sano-furukawa2008}. Since NQE do not seem to affect significantly the positions of the two peaks in the PCF, the distances for hydrogen and deuterium are expected to be similar, even though quantum effects are weaker for deuterium than for protons.

Upon compression, the O-O mean distance decreases (see Figure \ref{oo_dis}); this is correlated with a dilatation of the O-H bond and a concomitant contraction of the H$\cdots$O length. 
As pressure further increases, the two peaks merge until they become hardly distinguishable from one another; at high pressures, the ionocovalent and hydrogen bonds cannot be differentiated anymore. 
However, the PCF from the QTB simulations differ from those obtained via standard Langevin MD: the peaks are broader than their classical counterparts, and they merge at a lower pressure, even though the positions of the peaks are similar. 
Moreover, the minimum of the PCF between the two peaks is much higher when NQE are taken into account, suggesting that either disorder or proton tunneling is occuring at low pressure.
The PCF thus shows a merging of the two peaks, therefore indicating a structural change within the explored pressure range, however actual symmetrization does not occur: at $P=25$GPa, the oxygen-oxygen distance is larger than 2.4\AA\, (Figure \ref{oo_dis}), while the O-H distance at the same pressure is less than 1.1\AA, that is, less than half the O-O distance. This means that if a structural phase transition occurs, the PCF never becomes symmetric within the explored pressure range.

%%%%%%%%%%%%%%%%%%%%%%%%%%%%%%%%%%%%%
%  PROTON TUNNELING   %
%%%%%%%%%%%%%%%%%%%%%%%%%%%%%%%%%%%%%
\subsection{Proton tunneling}
\label{sec:hop}

The analysis of the PCF in section \ref{sec:pcf} does not allow to distinguish between the intrinsic width of the two peaks on the one hand and proton hopping on the other. 
The determination of a precise transition pressure from the computed distributions is therefore problematic.  
The usual relevant proton transfer coordinate in symmetrization transition is $\xi=\mathrm{d}_{\text{H}\cdots\text{O}} - \mathrm{d}_{\text{O-H}}$ where d$_{\text{H}\cdots\text{O}}$ and d$_{\text{O-H}}$ are the O-H distances between a given proton H and its two nearest-neighbor oxygen atoms \citep{benoit2005,morrone2009,lin2011}. 
Thus, when the proton is covalently bonded to one oxygen, we have $\xi>0$ while $\xi \approx 0$ indicates that the proton is located at the midpoint of the O-O segment.

\begin{figure}[tbp!]
\centering
\includegraphics[width=0.8\linewidth]{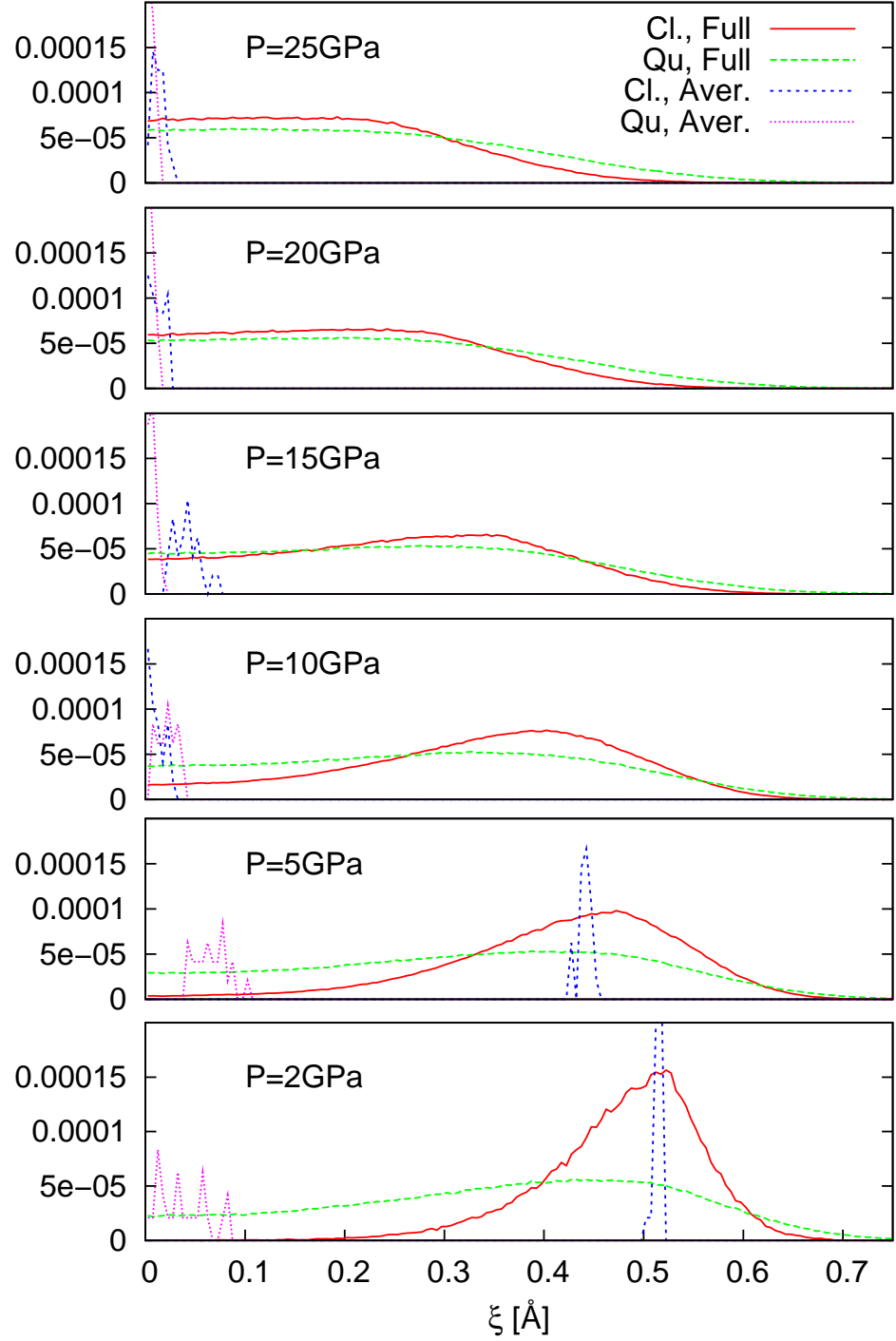}
\caption{Proton transfer coordinate distribution $p(\xi)$, $\xi = \mathrm{d}_{\text{H}\cdots\text{O}} - \mathrm{d}_{\text{O-H}}$ as a function of pressure in $\delta$-AlOOH. The distributions are obtained through the time-averaged atomic position (blue and magenta) and the instanteneous atomic positions (red and green). Results are shown for both the Quantum Thermal Bath (Qu.) simulation, and classical Langevin (Cl.) simulations.}
\label{xi}
\end{figure}

Figure \ref{xi} shows the distribution of the proton transfer coordinate computed in two different ways: $p(\langle\xi\rangle_t)$ from the time-averaged atomic positions and $p(\xi)$ from the instantaneous atomic positions. 
The time-averaged results show a transition to a symmetric state ($\langle\xi\rangle_t\approx 0$) at approximately 10 GPa, which is consistent with $Pnnm$ space group, while the full distribution is not peaked at $\xi=0$. 
This means that the average symmetry is that observed experimentally, but the system displays a large degree of disorder and the local instantaneous configurations may have lower symmetry than the average ones.
Moreover, at low pressure, the iono-covalent O-H bond is longer and the hydrogen bond shorter within QTB MD with respect to classical Langevin dynamics, in agreement with the results on the position of the peaks in the PCF (see Figure \ref{pcf}). 
We note also that our $\langle\xi\rangle_t$ from standard Langevin MD \emph{i.e.} with virtually classical protons, are in good agreement with experimental measurements on $\delta$-AlOOD \citep{sano-furukawa2009}, the deuterium nucleus being heavy enough to be treated as a classical particle. 
Moreover, we also compute the evolution of mean $\langle\xi\rangle_t$ with pressure in $\alpha$-AlOOH (diaspore), where no symmetrization occurs in the pressure range studied here (up to $30$ GPa), which is in agreement with experimental results \citep{winkler2001,friedrich2007}.
In addition, the distinction between results from QTB MD and standard Langevin MD is negligible: in diaspore, nuclear quantum effects do not play any major role in the hydrogen bond behavior, in contrast with the $\delta$ phase. 

A closer look to the full proton transfer coordinate distribution $p(\xi)$ that is computed from the instantaneous atomic positions allows to obtain more detailed information on the role of thermal and nuclear quantum effects along the symmetrization of the hydrogen bonds in $\delta$-AlOOH.
Within the classical frame, $p(\xi)$ shows a well defined maximum at $\xi_M$ for pressures up to $\simeq 15$ GPa, which shifts from $\xi_M\simeq 0.5$ {\AA} at $P=2$ GPa to $\xi_M\simeq 0.3$ {\AA} at $P=15$ GPa. 
At larger pressures, the proton transfer coordinate distribution becomes flat in a distance range which contracts when $P$ increases. 
Classical Langevin dynamics thus suggests a transition from a proton-ordered asymmetric hydrogen bond to a dynamically disordered symmetric hydrogen bond taking place around 15-20 GPa, where the variance of the distribution is largest. 
The picture that is obtained by the dynamics with the quantum thermal bath is in striking contrast with the classical one: the proton transfer coordinate distribution is quite flat even at low pressures, with $p(0)\neq 0$, which evidences the presence of a small but non null proton tunneling and a large variance - the maximum being $\xi_M\simeq 0.45$ {\AA} at $P=2$ GPa, close to its classical counterpart. 
The distribution flattens and its variance lessens for increasing pressures. 
Already at $P\simeq15$ GPa, the maximum $\xi_M\simeq 0.25$ is hardly distinguishable and $p(\xi_M) \simeq p(0)$. 

Therefore, according to the quantum frame, $\delta$-AlOOH undergoes a transition from a proton-disordered asymmetric hydrogen bond to a proton-disordered symmetric hydrogen bond. 
By direct comparison between the two frames, quantum effects enhance the proton disorder even at low pressures and slightly downshift the transition pressure with respect to purely thermal effects.  
The situation is thus somewhat complex: from X-ray scattering, in which proton disorder should be barely visible, only the averaged $Pnnm$ symmetry can be seen at high pressure. 
Proton tunneling at the time-scale of the experiment thus triggers the dynamical symmetrization of the hydrogen bond that leads to the $Pnnm$ hydrogen-centered (HC) phase, which is observed experimentally \citep{sano-furukawa2009,kuribayashi2014} for larger pressures than about 8 GPa. 
Its symmetrization from a classical viewpoint, that is when the potential felt by the protons only has one minimum, occurs at a higher pressure, around 30 GPa as found by Tsuchiya et al. \citep{tsuchiya2009}, as we will show in the next section that is devoted to the analysis of vibrational modes in $\delta$-AlOOH.

%%%%%%%%%%%%%%%%%%%%%%%%%%%%%%%%%%%%%%
% SECTION 3 : VIBRATIONAL PROPERTIES %
%%%%%%%%%%%%%%%%%%%%%%%%%%%%%%%%%%%%%%
\section{Vibrational properties}
\label{sec:spectra}
As previously pointed out, the proton distribution is difficult to establish through X-ray diffraction experiments, and pair-correlation functions as obtained from simulations are not always absolutely conclusive. 
On the other hand, vibrational spectroscopic measurements are an excellent means to detect experimentally phase transitions through mode softening \citep{aoki1996,goncharov1996,struzhkin1997,goncharov1999}, and it turns out these are also easily accessed through time-dependent simulations such as QTB molecular dynamics \citep{bronstein2014,brieuc2016}.

The vibrational spectra of $\delta$-AlOOH are rather complex \citep{tsuchiya2009}. 
Therefore, in order to determine which phonon modes are relevant to the hydrogen bond symmetrization in $\delta$-AlOOH, we conduct a preliminary mode analysis via the dynamical matrix, which is computed through density functional perturbation theory \citep{baroni2001} at $0$ K and without any quantum or anharmonic effects (section \ref{sec:prelim}), below and above the transition. Some of the relevant modes from the simulated vibrational spectra (at ambient temperature and including NQE) are then discussed in section \ref{sec:vibra}.

%%%%%%%%%%%%%%%%%%%%%%%%%%%%%%%%%%%%%%
% SECTION 3.1 : PRELIMINARY STUDY    %
%%%%%%%%%%%%%%%%%%%%%%%%%%%%%%%%%%%%%%
\subsection{Phonons: harmonic approximation at $T = 0$ K}
\label{sec:prelim}

In the asymmetric hydrogen bond configuration ($P2_1nm$, HOC) \citep{tsuchiya2008}, the vibrations at the Brillouin zone center $\Gamma$ consist in two high-frequency O-H stretching modes (around $2600$ cm$^{-1}$ at $0$ GPa), and four O-H$\cdots$O bending modes between $1100$ and $1350$ cm$^{-1}$.
The two low-frequency bending modes vibrate in the $\mathbf{c}$ direction, while the two bending modes at higher frequencies are in the $(\mathbf{a},\mathbf{b})$ plane, which contains the O-H$\cdots$O bonds.  When pressure increases, the O-H stretching mode frequencies decrease abruptly, while the bending mode frequencies increase slightly.

\begin{figure}[tbp!]
\centering
\includegraphics[width=\linewidth]{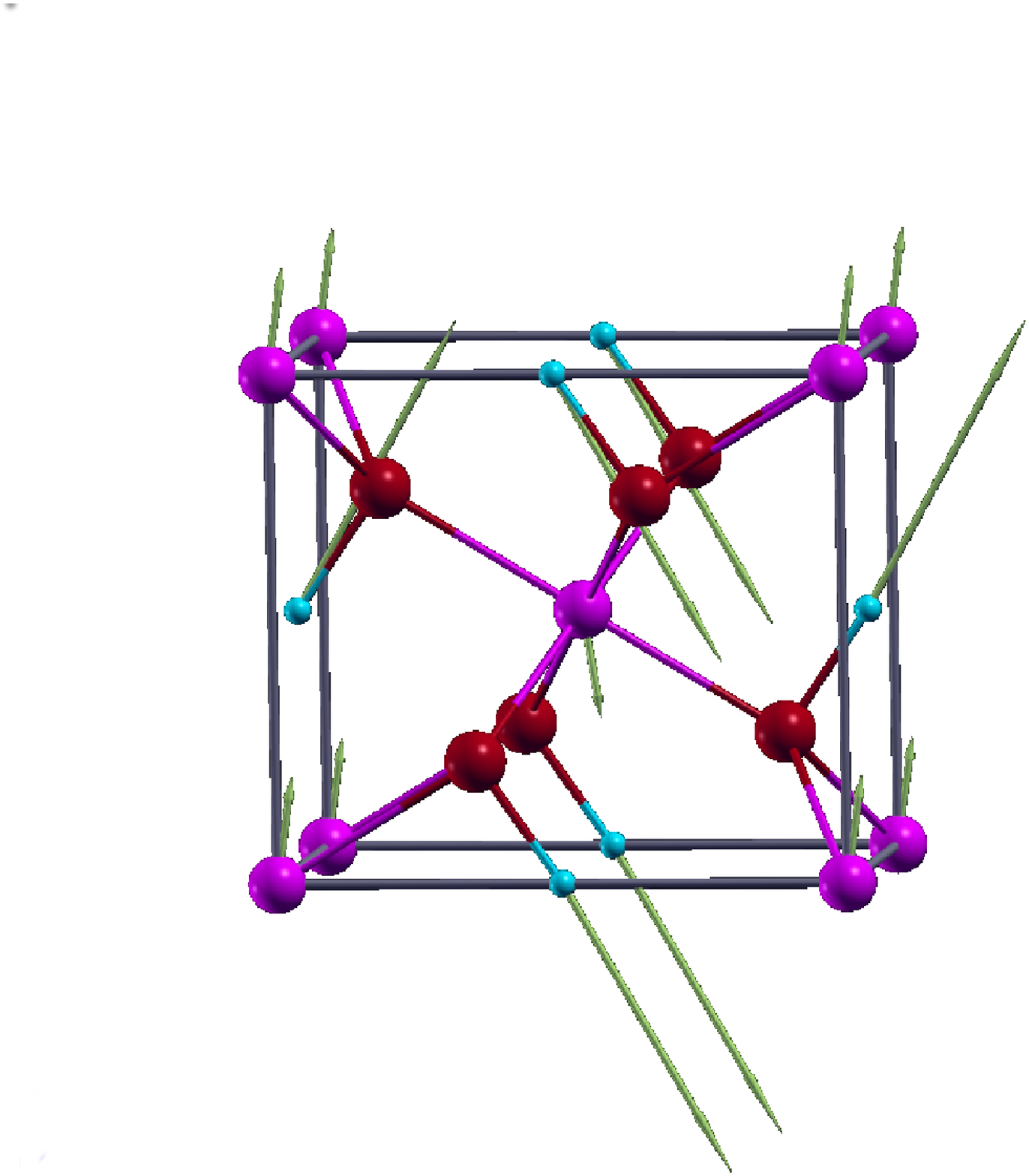}
\caption{Schematic representation of the atomic motion of $\delta$-AlOOH, for the imaginary frequency O-H stretching mode in the $Pnnm$ (HC) structure at $P=20$GPa. Color code: H in light blue, O in red, Al in violet. The crystallographic conventions used here are the same as in Reference \citep{tsuchiya2008}.}
\label{schema}
\end{figure}

In the symmetric phase ($Pnnm$, HC) \citep{tsuchiya2008} at $P=0$, the two highest O-H bending modes (around $1600$ cm$^{-1}$ at $\Gamma$) are in the $(\mathbf{a},\mathbf{b})$ plane and the two lowest ones along $\mathbf{c}$ vibrate in the $1300$ - $1400$ cm$^{-1}$ range. 
In contrast, the O-H stretching modes at $\Gamma$ have imaginary frequencies for pressures below 30 GPa (see figure \ref{schema}), their modulus tends to zero with increasing $P$.
This clearly indicates that the HC phase should be unstable below that pressure.
At 30 GPa, the stretching frequencies at the Brillouin zone center become real and then increase with pressure. 
At $P\simeq 50$ GPa they mix with bending modes, in the 1400 - 1700 cm$^{-1}$ range and harden with further increasing pressure.

As already shown by \cite{tsuchiya2002,tsuchiya2008}, this is consistent with a second-order transition from the HOC to the HC structure in $\delta$-AlOOH, which is accompanied by a strong softening of the O-H stretching modes. Therefore, our dynamical matrix analysis at $T=0$K and without NQE confirms a transition at $P\approx 30$ GPa, which, however is about 20GPa above the experimental transition pressure from \cite{kuribayashi2014}, a far from negligible difference. 
Moreover, the imaginary frequency O-H stretching mode also show motion of the Aluminium atoms (figure \ref{schema}) which means that the symmetry of the position thereof with respect to the hydrogen environment is unstable: this is reminiscent of the local symmetry as discussed in section \ref{sec:hop}. Finally, for some pressures around 30GPa, the bending and stretching modes can mix. Hence, we cannot simply rely on the description of the vibrations in $\delta$-AlOOH via the dynamical matrix approach close to the HOC - HC transition. 
We can indeed expect both thermal and quantum effects to allow the system to explore the anharmonic regions of phase space, as suggested by the issue of local versus average symmetric positions as done in section \ref{sec:hop} for protons. 
Thus, we turn to the study of the atomic trajectories in order to take into account the inherent anharmonicity of the system, as induced by thermal and quantum effects.

Nevertheless, the dynamical matrix approach allows us to choose a basis of eigenvectors - the phonon directions $\mathbf{u}_j$, $j \in [1, 5]$ - that is a key to the interpretation of the hydrogen vibrational spectra.
In the $P2_1nm$ configuration, $\mathbf{u}_1$ is parallel to the O-O direction, while $\mathbf{u}_2$ is orthogonal to $\mathbf{u}_1$ in the $(\mathbf{a},\mathbf{b})$ plane.
In the $Pnnm$ configuration, the O-H stretching mode eigenvector $\mathbf{u}_3$ forms a small angle with the O-O direction, and $\mathbf{u}_4$ is orthogonal to $\mathbf{u}_3$ in the $(\mathbf{a},\mathbf{b})$ plane.
In both configurations, $\mathbf{u}_5$ is parallel to the $\mathbf{c}$ direction.
The eigenvectors from the two phases are quite similar: in particular,  $\mathbf{u}_1 \approx \mathbf{u}_3 $ and $\mathbf{u}_2 \approx \mathbf{u}_4$.
Therefore, the use of these eigenvectors to obtain mode-specific density of states (see Equation \ref{eq:proj_dos}) appears to be reliable in both phases. 
At the transition, the OH modes can mix and become a linear combination of harmonic ones; apart from the modes here reported, there are no others involving proton displacements in the harmonic case. 
Therefore, the $\mathbf{u}_j$ vectors form a minimal basis set to analyze the behavior with pressure of modes involving protons.

%%%%%%%%%%%%%%%%%%%%%%%%%%%%%%%%%%%%%%%%%%%%%%%%%
% SECTION 3.2 : MOLECULAR DYNAMICS TRAJECTORIES %
%%%%%%%%%%%%%%%%%%%%%%%%%%%%%%%%%%%%%%%%%%%%%%%%%
\subsection{Vibrational spectra from molecular dynamics trajectories}
\label{sec:vibra}
\begin{figure}[tbp!]
\centering
\includegraphics[width=\linewidth]{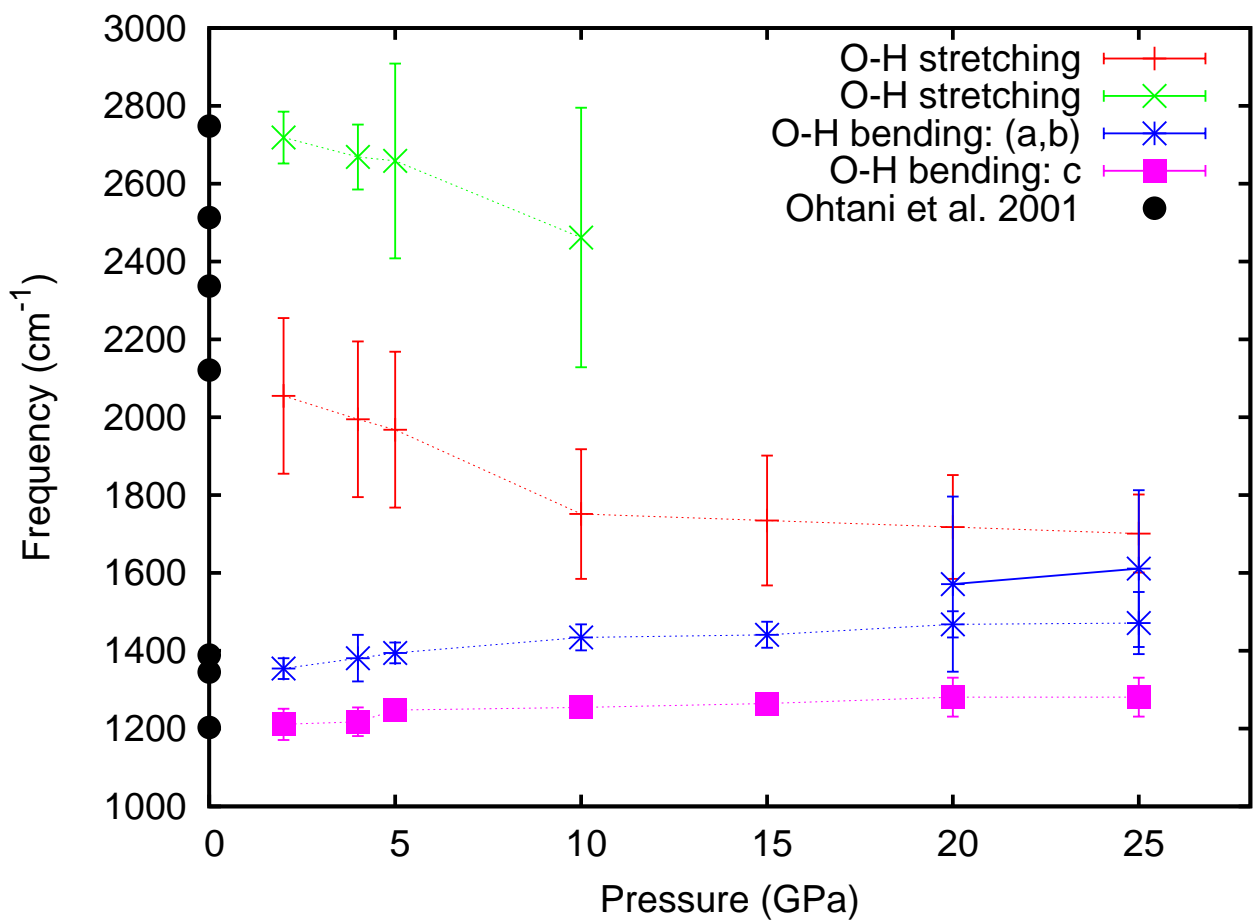}
\caption{Pressure dependence of the hydrogen vibration modes frequencies in $\delta$-AlOOH, obtained from Quantum Thermal Bath molecular dynamics simulations at ambient temperature. Upon compression, the O-H stretching modes soften drastically (red and green curves), while the O-H bending modes frequencies (in the $(\mathbf{a},\mathbf{b})$ plane in blue, and in the $\mathbf{c}$ direction in purple) slightly increase. The black dots indicate Raman measurements on $\delta$-AlOOH.\citep{ohtani2001} The vertical bars indicate the width of the peaks obtained via the Fourier Transform.}
\label{spec}
\end{figure}

 In order to take into account both thermal and quantum effects, as well as the anharmonicities of the system, the vibrational spectrum of hydrogen in $\delta$-AlOOH is calculated directly from our MD simulations, through the Fourier Transform of the autocorrelation function involving the different nuclei \citep{marshall1971}.
For each pressure, the hydrogen spectrum density $I(\omega)$ is proportional to: 
\begin{equation}
I(\omega)\sim \sum_k \left\vert \int d_\mathrm{OH}(t) \mathrm{e}^{i\omega t} \mathrm{d}t \right\vert^2
\end{equation}
where $d_\mathrm{OH}$ is the shortest O-H distance (corresponding to the O-H ionocovalent bond) and the sum runs over all H atoms. As $I(\omega)$ contains information about all hydrogen vibrations, we also compute the partial spectrum densities $I_j(\omega)$ in order to unravel the distinct mode contributions: 
\begin{equation}
I_j(\omega) \propto \sum_k \left| \int \chi_k^{(j)} (t) \mathrm{e}^{i\omega t} \mathrm{d}t \right|^2, \quad j=1, \dots 5
\label{eq:proj_dos}
\end{equation}
where $ \chi_k^{(j)}$ is the projection of the hydrogen position vector $\mathbf{r}_k$ onto one of the vectors $\mathbf{u}_j$ defined in the previous section. 
Each partial spectrum $I_j(\omega)$ then corresponds to hydrogen vibrations along the eigenvector $\mathbf{u}_j$. 
By comparing the full spectrum $I(\omega)$ to the peaks of $I_j(\omega)$, we determine the frequencies of the distinct hydrogen modes: the O-H stretching modes and the O-H$\cdots$O bending modes. 
The evolution of these frequencies with increasing pressure is shown in Figure \ref{spec}, as well as the width of the corresponding peaks (indicated by the vertical bars). 

At low pressure, we distinguish two stretching modes at approximately $2800$ and $2100$ cm$^{-1}$ and two bending modes around $1400$ cm$^{-1}$ (in the $(\mathbf{a},\mathbf{b})$ plane) and $1200$ cm$^{-1}$ (in the $\mathbf{c}$ direction). 
These results are in very good agreement with Raman measurements \citep{ohtani2001} at ambient pressure. 
Due to the large widths of the peaks in the calculated spectrum, we could distinguish only two high-frequency O-H stretching modes, while Ohtani and coworkers observed four broad bands \citep{ohtani2001}. 
Tsuchiya and coworkers suggested that the presence of these multiple bands, instead of two sharp peaks as calculated in the dynamical matrix approach (see section \ref{sec:prelim}), is due to hydrogen disorder at low pressure \citep{tsuchiya2008}, which is consistent with our results on the O-H pair correlation function. 
Upon compression, the O-H stretching modes soften gradually, their frequencies dropping to approximately $1800$ and $2500$ cm$^{-1}$. 
Moreover, the width of the high-frequency mode increases, which implies that the two peaks of $I(\omega)$ merge at the transition and are no longer distinguishable. 
In contrast, the bending mode frequencies increase slightly with pressure. Above a critical pressure ($P_c = 10$ GPa), the O-H stretching mode frequency ($\simeq 1800$ cm$^{-1}$) does not vary appreciably with pressure, at least up to $25$ GPa, while the bending mode frequencies increase continuously. 
The second O-H bending mode in the $(\mathbf{a},\mathbf{b})$ plane becomes distinguishable only above $20$ GPa. 
The previous trends agree with those found by Tsuchiya and coworkers \citep{tsuchiya2008}.

The softening of the O-H stretching mode up to about $10$ GPa, where bending and stretching modes mix up, is consistent with a second-order phase transition as also guessed by the discontinuity of the compressibility.
This is also quite close to the high-pressure behavior of the O-H vibrations during the transition from asymmetric phase VII to symmetric phase X in ice \citep{aoki1996,goncharov1996,struzhkin1997,goncharov1999,bronstein2014,bronstein2016}.
Above this critical pressure, there is a continuous hardening of the O-H-O bending modes, while the stretching mode frequencies do not vary in our pressure range (up to $25$ GPa). 
By extrapolating the trends shown in Figure \ref{spec}, it appears that the bending modes show higher frequencies than the stretching modes above $\sim 30$ GPa, as also predicted by the dynamical matrix analysis in section \ref{sec:prelim}. 

All the previous observations conspire to indicate that $P=10$ GPa is close to the critical pressure $P_c$ for the transition from a proton-disordered 2$_1$nm phase with asymmetric hydrogen bonds to a proton-disordered $Pnnm$ phase with symmetric hydrogen bonds. 
The proton disorder decreases with pressure and a proton-ordered  $Pnnm$ phase with symmetric hydrogen bonds is recovered at pressures as high as $\simeq 30$ GPa.

%%%%%%%%%%%%%%%%%%%%%%%%%%%%
% SECTION 5 : CONCLUSION   %
%%%%%%%%%%%%%%%%%%%%%%%%%%%%
\section{Conclusion}
We study the symmetrization of the Hydrogen bond in the high-pressure $\delta$ phase of AlOOH by molecular dynamics simulations, thus including thermal effects. For the first time to our knowledge, we also include nuclear quantum effects in the description of the transition through a recent though relatively simple method - the Quantum Thermal Bath (QTB) - in which semi-classical atomic trajectories fulfilling the Bose-Einstein statistics are generated via the contact with a stochastic bath of quantum harmonic oscillators. From a detailed comparison between three simulation frames (static calculations, classical molecular dynamics and quantum molecular dynamics simulations) we disentangle thermal and quantum effects.
 
The transition can be schematically depicted as proceeding from a low-pressure hydrogen off-centered (HOC) phase with $P2_1nm$ space group to a more symmetric, hydrogen-centered (HC) phase with $Pnnm$ space group at high pressure. The HC-$Pnnm$ phase is harder than the HOC-$P2_1nm$ one, with important consequences on the stability and mechanical properties of $\delta$-AlOOH in the lower Earth mantle.

The static calculations fully agree with previous ones \citep{tsuchiya2002, Li2006, tsuchiya2008} indicating that a pressure as high as $\simeq 30$ GPa is needed to stabilize the proton-ordered HC-$Pnnm$ phase. 
The apparent contradiction with recent X-ray diffraction data \citep{kuribayashi2014} providing experimental evidence that the transition to the high-symmetry phase happens at much lower pressures, around 8 GPa, is solved by analyzing the role of thermal and nuclear quantum effects in the transition. 
The onset of proton tunneling appears at rather low pressures, below 10 GPa, which is much lower than that needed to shorten O-O distances down to $\simeq 2.42$ {\AA} - the usual threshold for the hydrogen-bond symmetrization \citep{benoit2005,morrone2009,lin2011}. 
Proton tunneling implies that the time-averaged proton site coincides with the O-O midpoint, although at $P<15-20$ GPa this is still not an equilibrium position, neither in the classical nor in the quantum frames. 
However, the average structure fully coincides with the HC-$Pnnm$ phase. Any structural measurement that is recorded on larger time lapses than the proton tunneling time would therefore provides such a picture. 
Nevertheless, only at larger pressures (about 20 GPa when NQE are included or around 25 GPa in classical simulations, at room temperature) the symmetric hydrogen bond corresponds to a maximum of the proton distribution. 
We propose that the measurement of the Debye-Waller factors through quasi-elastic neutron scattering \citep{teixeira1999} could distinguish between the two regimes.

As in other order-disorder phase transitions, the analysis of the O-H modes is crucial to point out the transition. 
From the trends of H-related modes with pressure, which we extracted from the dynamical simulations including NQE as well as anharmonic effects, the transition should happen at pressures as low as 10 GPa, where the O-H stretching modes considerably weaken and eventually fade out. 
Those trends are fully consistent with the analysis of the structural observables (average positions and pair-correlation functions) but they yield a much clearer signature of the transition as phonon modes are directly linked with the change of character of the proton dynamics. 
We suggest that high-resolution vibrational spectroscopy could complement the present knowledge and contribute to better identify the precise transition pressure. In order to point out the role of nuclear quantum effects, the experiments could be ideally conducted on both AlOOD and AlOOH. 

Our results might also be relevant for other hydrous minerals, such as guyanaite \citep{jahn2012} or dense hydrous magnesium silicates \citep{ohira2014,tsuchiya2013}, where nuclear quantum effects, such as tunneling, could be at work.

%%%%%%%%%%%%%%%%%%%%%%%%%%%%%%%%%%%%%%%%%%%%%%%%%%%%%%%%%%%%%%%%%%%%%%
% --------------------------------------------------------------------
%              ACKNOWLEDGEMENTS
% --------------------------------------------------------------------
%%%%%%%%%%%%%%%%%%%%%%%%%%%%%%%%%%%%%%%%%%%%%%%%%%%%%%%%%%%%%%%%%%%%%%

%\begin{acknowledgments}
\vspace{1cm}\textbf{Acknowledgments:}\\
We thank E. Balan for useful discussions and a critical reading of the manuscript. 
Y.B. acknowledges financial support from the Conseil R\'egional d'\^Ile-de-France (DIM Oxymore).
This work was granted access to the HPC resources of CINES under the allocation 2016096719 made by GENCI.
%\end{acknowledgments}

%%%%%%%%%%%%%%%%%%%%%%%%%%%%%%%%%%%%%%%%%%%%%%%%%%%%%%%%%%%%%%%%%%%%%%
% --------------------------------------------------------------------
%               BIBLIOGRAPHY
% --------------------------------------------------------------------
%%%%%%%%%%%%%%%%%%%%%%%%%%%%%%%%%%%%%%%%%%%%%%%%%%%%%%%%%%%%%%%%%%%%%%

% --------------------------------------------------------------------
%                     FIGURES
% --------------------------------------------------------------------

%%%%%%%%% FIGURE 1 %%%%%%%%%%%

%%%%%%%% FIGURE 2 %%%%%%%%

%%%%%%%% FIGURE 3 %%%%%%%%%%

%%%%%%%% FIGURE 4 %%%%%%%%%%

%%%%%%% FIGURE 5 %%%%%%%%%%%%%

%%%%%%%%%%%% FIGURE 6 %%%%%%
\end{document}